\documentclass[11pt,a4paper]{article}

\usepackage[utf8]{inputenc}
\usepackage[T1]{fontenc}
\usepackage{amsmath,amssymb,amsfonts}
\usepackage{graphicx}
\usepackage{booktabs}
\usepackage{float}
\usepackage{hyperref}
\usepackage{xcolor}
\usepackage[margin=1in]{geometry}

\hypersetup{
    colorlinks=true,
    linkcolor=blue,
    citecolor=blue,
    urlcolor=blue
}

\newcommand{\nrmse}{\mathrm{NRMSE}}

\title{When Classical Baselines Are Tuned as Carefully\\
as the Quantum Model, Does Quantum Reservoir\\
Computing Still Win?}

\author{
    Tushar Pandey\\
    \textit{Texas A\&M University}
}

\date{\today}

\begin{document}

\maketitle

\begin{abstract}
Can a small quantum computer forecast a changing signal better than an ordinary
classical method? Many studies say yes, but the classical methods they compare
against are often left in a basic, untuned state while the quantum model is
carefully optimised. We ask what happens when the classical competitor is given
exactly the same care: the same size and the same amount of tuning effort. We
study two popular reasons a quantum reservoir is thought to help, using exact
simulations of small quantum systems (up to eleven qubits) on prediction tasks.
In both cases the quantum advantage disappears once the comparison is fair. In
the first, extra quantum measurements add nothing that a simple classical formula
of the same size does not already provide. In the second, a feedback loop genuinely
helps the quantum model, turning a useless setup into a working predictor, yet a
well-tuned classical network still predicts slightly more accurately, and the gap
is statistically reliable. Our point is not that quantum reservoirs can never win,
but that two of their commonly cited advantages do not hold up against fair
classical competitors at this scale. We provide these matched comparisons as a
simple, reusable checklist for honest benchmarking. All results are fully
reproducible from fixed random seeds.
\end{abstract}

\section{Introduction}

Reservoir computing~\cite{jaeger2001,maass2002} trains only a linear readout on
top of a fixed nonlinear dynamical system, which makes it an attractive
near-term application of quantum hardware: a fixed quantum reservoir avoids the
trainability pathologies of variational circuits~\cite{mcclean2018} while
supplying a high-dimensional nonlinear feature map. Since the original
proposals~\cite{fujii2017,nakajima2019}, a growing body of work reports that
quantum reservoir computing (QRC) outperforms classical reservoirs and recurrent
networks on chaotic, financial, and biomedical time
series~\cite{mujal2021,martinezpena2021}. At the same time, dequantization and
expressivity results~\cite{schuld2021,sweke2021} caution that the function
classes realised by such circuits can often be reproduced classically, which
makes the choice of classical baseline decisive.

A recurring weakness in these reports is the classical baseline. When a quantum
model is compared against an echo-state network (ESN)~\cite{jaeger2001} or a
polynomial readout,
the strength of the conclusion depends entirely on whether the classical model
was given an equal chance: the same feature-space capacity, the same
hyperparameter-tuning budget, the same causal evaluation protocol, and a metric
that does not implicitly favour the quantum side. In practice these conditions
are often not met. Baselines are reported with default settings while the
quantum model is tuned; ``classical reservoir'' refers to a different
architecture than the quantum one; efficiency is measured as simulation
wall-clock time rather than any hardware-relevant quantity; and directional
accuracy is reported without the persistence control that makes it meaningful.

This paper asks a narrow question: \emph{when the classical
baseline is matched in capacity and tuning, does the quantum reservoir's
reported advantage survive?} We study two representative advantage mechanisms in
noiseless statevector simulation at small system size ($q\le 11$) and find that
on these tasks it does not. Our contribution is not the negative
outcome per se but the two controls that produce it: a capacity-matched
classical model and a tuning-matched classical recurrent network, together with
a correctness-gated, deterministic simulation harness. We state the scope
plainly: this is a simulation study at sizes that classical hardware can
represent exactly ($q \le 11$); the claim is ``no advantage here, under fair
comparison,'' not ``no advantage is possible.''

We focus on these two advantage mechanisms because they admit the cleanest
matched controls. They are part of a broader fair-comparison cartography of QRC
advantage claims: a capacity-matched scaling study of the same reservoir
architecture appears in a companion paper~\cite{pandey2026qra}, and additional
angles (high-body trajectory observables, channel-spectrum learning, and
cross-asset volatility spillover) are deferred to future work; the methodology
and the two controls developed here apply to those settings as well.

\section{Methods}
\label{sec:methods}

\subsection{Common harness and fairness controls}

All experiments share a single evaluation harness with the following
properties, several of which are the fairness controls whose absence we are
critiquing.

\paragraph{Causal, leakage-free evaluation.} Each series is split
chronologically into training, validation, and test segments. Every
preprocessing step (standardisation) and every hyperparameter is fit on the
training/validation portion only and applied causally; the test segment is
touched once, for the final reported number.

\paragraph{Validation-only tuning, applied equally.} All free
hyperparameters (ridge penalty for every method, quantum evolution time
$\tau$, feedback gain $k_{\mathrm{fb}}$, and the ESN spectral radius and leak
rate) are selected by validation error. Importantly, the classical baselines
receive the \emph{same} tuning budget as the quantum model. This is the single
most important control and the one most often missing.

\paragraph{Matched feature dimension.} Where a quantum and a classical reservoir
are compared, their readout feature dimensions are matched so that the
comparison is not confounded by readout width.

\paragraph{Correctness gates.} Before any result is generated, an automated
self-test verifies: the reservoir state is physical ($\mathrm{Tr}\,\rho=1$ and
all measured expectations lie in $[-1,1]$); the two-point readout computed from
the density-matrix diagonal agrees with the explicit operator expectation to
machine precision; finite-shot estimates converge to the exact values as the
shot count grows; and, for the feedback model, that the feedback path is wired
correctly (with gain zero it reproduces the open-loop reservoir to machine
precision) and is strictly causal (perturbing a future input leaves earlier
features unchanged). Every reported number is read back from disk and checked
against a content hash; all results in this paper are deterministic and
reproduce bit-for-bit.

\paragraph{Metrics.} We report normalised root-mean-square error
($\nrmse$, normalised by the per-target standard deviation, so that
$\nrmse=1$ is the trivial mean predictor) and, for directional tasks, mean
directional accuracy (MDA). We define MDA as the fraction of test points on
which the predicted direction of the next move matches the observed direction,
$\mathrm{MDA}=\frac{1}{|\mathrm{test}|}\sum_t \mathbf{1}\!\left[\operatorname{sign}(\hat y_{t+h}-y_t)=\operatorname{sign}(y_{t+h}-y_t)\right]$,
where $y_t$ is the last observed level. We report it alongside the
\emph{persistence-MDA}: the same directional accuracy achieved by the naive
predictor that assumes the most recent observed move continues
($\operatorname{sign}(y_t-y_{t-h})$). Persistence-MDA is the meaningful floor
for a directional task, since an MDA above $0.5$ is otherwise easy to obtain on
trended data.

\subsection{Quantum reservoirs}

The quantum reservoir is a fully connected transverse-field Ising system,
\begin{equation}
H = \sum_{i<j} J_{ij}\, X_i X_j + v \sum_i Z_i,
\qquad J_{ij}\sim\mathcal{U}[0,1],\ v=1,
\end{equation}
fixed after random initialisation. Classical inputs are angle-encoded via
single-qubit $R_Y$ rotations; the state evolves for time $\tau$; and the readout
consists of single-qubit expectations $\langle Z_i\rangle$ and, where indicated,
two-point correlators $\langle Z_i Z_j\rangle$, which are estimable from the same
computational-basis measurement record~\cite{huang2020}. A recurrent variant
carries memory across timesteps by partial-tracing the input qubits between
encodings, following standard recurrent-QRC constructions~\cite{nakajima2019}. Because $Z_i$ and $Z_iZ_j$
are diagonal in the computational basis, all correlators are obtained from the
state-diagonal in a single contraction, verified equal to the explicit operator
expectation (Section~\ref{sec:methods}).

\section{Case study I: a correlator readout adds nothing beyond a
capacity-matched polynomial}
\label{sec:corr}

\subsection{Setup}

A frequently proposed route to quantum expressivity is to enrich the readout
with two-point correlators $\langle Z_i Z_j\rangle$, which capture pairwise
quantum correlations and grow the feature space quadratically. The natural
question a fair comparison must ask is whether these features carry information
that an explicit classical quadratic model does not already have.

We test this on a controlled testbed of $N=5$ coupled H\'enon maps: bounded,
genuinely quadratic, multivariate dynamics with a tunable cross-asset coupling
$c$. We compare three models on one-step prediction (all with the same windowed
ridge readout and causal protocol): the tuned correlator QRC ($q=8$, readout
tuned over $\tau$, encoding scale, and single- vs.\ two-reservoir ensemble); a
degree-2 polynomial model (\textsc{Poly2}) whose features are all pairwise
products of the windowed inputs, the \emph{capacity-matched} classical
control, given \emph{more} features than the QRC; and the
concatenation $\textsc{Poly2}\oplus\textsc{QRC}$, which tests directly whether
the quantum features add anything to the polynomial basis. Observation noise of
standard deviation $0.1$ is added so that no model can solve the map exactly
through its generating polynomial. The concatenation ablation, running the
classical and quantum feature sets together and asking whether the union beats
the classical part alone, is exactly the control omitted by typical
hybrid-architecture reports.

\subsection{Results}

\begin{figure}[H]
\centering
\includegraphics[width=0.72\linewidth]{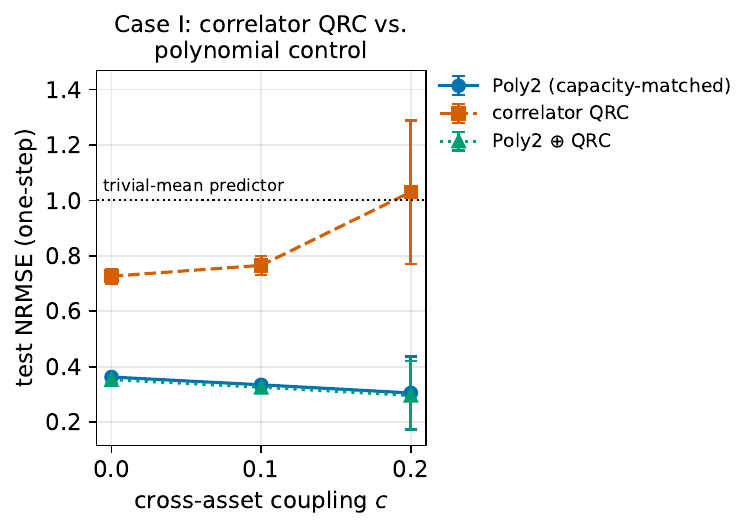}
\caption{Case study I. One-step test NRMSE versus cross-asset coupling $c$ for
the correlator QRC, the capacity-matched degree-2 polynomial (\textsc{Poly2}),
and their concatenation $\textsc{Poly2}\oplus\textsc{QRC}$. The polynomial
control is below the QRC at every coupling, and the concatenation tracks
\textsc{Poly2} almost exactly, so adding the quantum features changes nothing of
substance. The dotted line marks the trivial-mean predictor ($\nrmse=1$).}
\label{fig:caseI}
\end{figure}

Table~\ref{tab:corr} reports the verified numbers. At the one-step horizon the
\textsc{Poly2} control beats the tuned QRC outright at every coupling (e.g.\
$\nrmse$ $0.335$ vs.\ $0.766$ at $c=0.1$), and at the strongest coupling the
tuned QRC degrades to the trivial-mean predictor ($\nrmse=1.03$) while the
classical models remain predictive. Concatenating the quantum
features onto \textsc{Poly2} improves on \textsc{Poly2} alone by only
$\approx\!2.6\%$ at every coupling, the magnitude expected from a few extra mildly-regularising
features, not from complementary signal. The same verdict holds on mean
directional accuracy: \textsc{Poly2} reaches $\mathrm{MDA}\approx 0.96$
(persistence $\le 0.18$), the quantum model is no better, and the concatenation
does not exceed \textsc{Poly2}. The coupling axis, which one might expect to
favour the joint quantum encoding, instead makes the QRC relatively worse
(Fig.~\ref{fig:caseI}).

\begin{table}[H]
\centering
\caption{Case study I (one-step, observation noise $0.1$): correlator QRC vs.\
the capacity-matched \textsc{Poly2} control and their concatenation, on coupled
H\'enon dynamics at three couplings $c$. Lower $\nrmse$ is better;
$\nrmse<1$ is predictive. ``cat'' is $\textsc{Poly2}\oplus\textsc{QRC}$; the
final column is the relative $\nrmse$ reduction of cat over \textsc{Poly2}
(positive $=$ cat lower). Windowed feature dimensions: QRC $72$, \textsc{Poly2}
$135$, so the polynomial is given \emph{more} capacity than the quantum readout,
not less. Mean $\pm$ s.d.\ over five data$\times$reservoir seed pairs; all values verified from disk.}
\label{tab:corr}
\begin{tabular}{lcccc}
\toprule
$c$ & QRC & \textsc{Poly2} & cat & cat vs.\ \textsc{Poly2} \\
\midrule
$0.0$ & $0.726\pm0.026$ & $0.363\pm0.012$ & $0.354\pm0.012$ & $+2.6\%\pm0.3$ \\
$0.1$ & $0.766\pm0.034$ & $0.335\pm0.012$ & $0.326\pm0.007$ & $+2.6\%\pm1.5$ \\
$0.2$ & $1.030\pm0.258$ & $0.305\pm0.132$ & $0.297\pm0.125$ & $+2.7\%\pm1.0$ \\
\bottomrule
\end{tabular}
\end{table}

We are precise about what this does and does not show. The coupled
H\'enon dynamics are exactly quadratic, so a degree-2 polynomial spans the true
generating function class; this is by construction the most favourable possible
setting for \textsc{Poly2}, and the result is correspondingly bounded. The claim
we make is therefore not ``a fair baseline always erases the edge,'' but the
narrower and fully supported one: \emph{when the classical model already covers
the data's function class, a two-point correlator readout contributes nothing
beyond it}. The concatenation $\textsc{Poly2}\oplus\textsc{QRC}$ improves on
\textsc{Poly2} by only $\approx\!2.6\%$ (mean over five seeds, consistent across
all three couplings), the magnitude of mild regularisation rather
than of complementary signal. The correlator readout does help the QRC relative
to a single-point readout, but that gain does not transfer into an advantage
over a classical model of matched (indeed larger) capacity. A reader should read
Case~I as a clean refutation of the specific ``correlators add quantum
expressivity'' argument on quadratic data, not as a universal statement.

\section{Case study II: feedback is a live mechanism but loses to a
tuning-matched ESN}
\label{sec:fb}

\subsection{Setup}

A distinct advantage claim concerns \emph{feedback-driven} QRC, in which the
measurement outcome at one step modulates the reservoir drive at the next,
making the effective dynamics path-dependent rather than a fixed function of a
finite input window. Path-dependence is something a fixed-window polynomial
cannot reproduce, but it is precisely what a classical recurrent network also
provides. The honest baseline for a feedback reservoir is therefore not a
polynomial but a recurrent network, the echo-state network (ESN), and the test
is whether \emph{quantum} recurrence beats \emph{classical} recurrence at equal
tuning.

We use a nonstationary task built so that recurrence is necessary: a hidden
two-state Markov process switches the sign of a first-order autoregressive
coefficient, so each regime is individually predictable but the regime must be
inferred online from the observable history. A correctness gate confirms the
task genuinely requires regime inference (an oracle given the hidden state beats
a fixed-window linear model, with the gap growing as the switching rate rises).
We implement feedback as an additive, saturating term on the input drive angle,
$\phi_t = x_t + k_{\mathrm{fb}}\tanh(\bar m_{t-1})$, where $\bar m_{t-1}$ is the
mean measurement at the previous step; $k_{\mathrm{fb}}=0$ recovers the
open-loop reservoir exactly. We compare, over ten data seeds $\times$ ten
reservoir seeds ($100$ paired runs; an initial $3\times3$ grid served only as
a pilot), the tuned feedback QRC, its open-loop counterpart, a
tuning-matched ESN (spectral radius and leak rate selected on the same
validation budget), \textsc{Poly2}, and a linear model.

\subsection{Results}

Feedback is a genuine, live mechanism. In the $10\times10$-seed comparison at
$p_{\mathrm{sw}}=0.05$ (Table~\ref{tab:fb}, lower), the open-loop reservoir is
useless ($\nrmse = 1.020 \pm 0.036$, worse than the mean predictor) while
closing the loop lifts it to $0.787 \pm 0.059$, a $23\%$ improvement on
identical seeds. A single-seed switching-rate sweep (Table~\ref{tab:fb}, upper)
shows the same effect across $p_{\mathrm{sw}}\in\{0.02,0.05,0.10\}$
($+10$ to $+19\%$); we report it as an illustrative trend, with the load-bearing
seed-averaged number taken from the lower block. This is, to our knowledge, the
clearest case in our study of a quantum reservoir mechanism doing something its
open-loop version cannot.

\begin{table}[H]
\centering
\caption{Case study II. Upper: feedback vs.\ open-loop QRC across switching
rates $p_{\mathrm{sw}}$ (single seed, illustrative). Lower: the fair test at
$p_{\mathrm{sw}}=0.05$ over $10\times10$ seeds (mean $\pm$ s.d.), one-step
$\nrmse$ and MDA (persistence MDA $=0.27$). The tuning-matched ESN beats the
feedback QRC; the paired NRMSE difference is significant
($t$-test $p\approx10^{-21}$, Wilcoxon $p\approx10^{-16}$, bootstrap $95\%$
CI $[0.057,0.078]$).}
\label{tab:fb}
\begin{tabular}{lccc}
\toprule
\multicolumn{4}{l}{\emph{Upper: feedback vs.\ open-loop (single seed)}}\\
$p_{\mathrm{sw}}$ & open-loop & feedback & gain \\
\midrule
$0.02$ & $1.007$ & $0.874$ & $+13.2\%$ \\
$0.05$ & $1.018$ & $0.824$ & $+19.1\%$ \\
$0.10$ & $1.032$ & $0.925$ & $+10.5\%$ \\
\midrule
\multicolumn{4}{l}{\emph{Lower: fair comparison, $10\times10$ seeds}}\\
method & $\nrmse$ & MDA & \\
\midrule
ESN (tuned)      & $0.720\pm0.060$ & $0.737\pm0.054$ & \\
Linear           & $0.750\pm0.056$ & $0.735\pm0.055$ & \\
\textsc{Poly2}   & $0.762\pm0.046$ & $0.734\pm0.055$ & \\
Feedback QRC     & $0.787\pm0.059$ & $0.728\pm0.051$ & \\
Open-loop QRC    & $1.020\pm0.036$ & $0.740\pm0.058$ & \\
\bottomrule
\end{tabular}
\end{table}

\begin{figure}[H]
\centering
\includegraphics[width=0.72\linewidth]{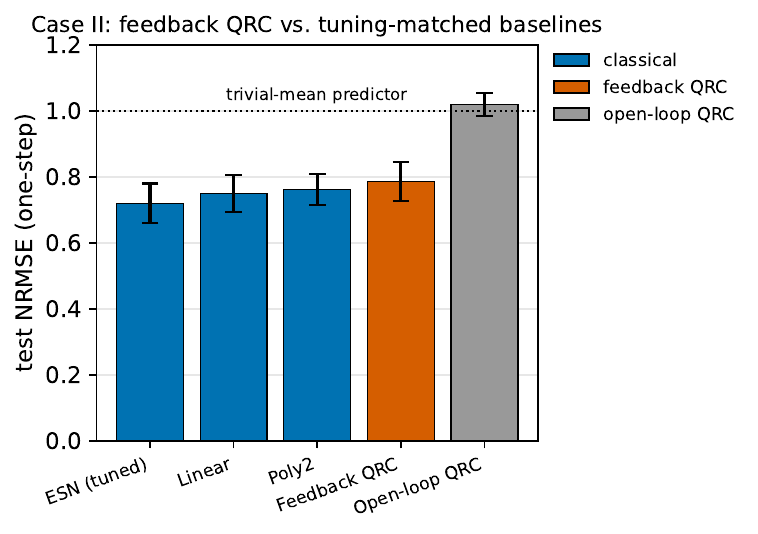}
\caption{Case study II. One-step test NRMSE at $p_{\mathrm{sw}}=0.05$
(mean $\pm$ s.d.\ over $10\times10$ seeds). Closing the feedback loop rescues the
useless open-loop reservoir (rightmost) to predictive performance, but the
tuning-matched ESN and even a plain linear model remain ahead of the feedback
QRC. Blue bars are classical; the dotted line is the trivial-mean predictor.}
\label{fig:caseII}
\end{figure}

The mechanism being live, however, does not make it advantageous. Under the fair
$10\times10$-seed comparison (Table~\ref{tab:fb}, lower), the tuning-matched ESN
attains $\nrmse = 0.720 \pm 0.060$ against the feedback QRC's
$0.787 \pm 0.059$, an ESN advantage of $8.5\%$. Because the $100$ runs are paired
(shared data and reservoir seeds), we test the per-run difference directly: the
mean paired difference is $0.067$ in NRMSE (ESN better), overwhelmingly
significant under both a paired $t$-test ($t=12.3$, $p\approx1.4\times10^{-21}$)
and the Wilcoxon signed-rank test ($p\approx9\times10^{-17}$), with a bootstrap
$95\%$ confidence interval of $[0.057,\,0.078]$ that excludes zero. All three
of these tests operate \emph{across} the $100$ seeds (the across-Hamiltonian
distribution): the paired $t$- and Wilcoxon tests take the $100$ per-run
ESN-minus-QRC NRMSE differences as their sample, and the bootstrap resamples
those same $100$ paired per-run differences with replacement ($20{,}000$
resamples), so the interval reflects across-seed/Hamiltonian variability rather
than within-run test-point noise. The $92\%$ figure below is likewise the
fraction of the $100$ paired runs on which the ESN wins. The ESN beats
the feedback QRC in $92\%$ of the paired runs, so the result is systematic rather
than driven by a few realisations; and the feedback QRC's spread across the $100$
reservoir/data realisations ($\mathrm{s.d.}=0.059$) is no larger than the ESN's
($0.060$), confirming the quantum reservoir is not merely an unlucky draw. The
feedback QRC in fact trails
every classical baseline, including a plain windowed linear model
($0.750$); note that on this task the quadratic \textsc{Poly2} features ($0.762$)
are themselves worse than the linear model, a reminder that added capacity is not
automatically useful. On directional accuracy all models cluster near $0.73$, far
above the $0.27$ persistence floor but statistically indistinguishable across
methods
(Fig.~\ref{fig:caseII}). The conclusion is
that quantum feedback here reconstructs the recurrent memory that a classical
recurrent network already provides, and does so slightly less efficiently;
closing the loop rescues the quantum reservoir up to, but not past, the
classical pack.

\section{Discussion}

Two mechanisms, two representative advantage claims, one outcome: under a
capacity-matched classical control in the first case and a tuning-matched
classical recurrent control in the second, the quantum reservoir's reported edge
does not survive. We emphasise three points.

First, the controls are the contribution. The concatenation ablation
($\textsc{Poly2}\oplus\textsc{QRC}$ vs.\ \textsc{Poly2}) directly measures
whether quantum features add information to a classical model of matched
capacity, and the same-budget ESN measures whether quantum recurrence beats
classical recurrence on equal footing. These are inexpensive to run and clear
in interpretation, yet they are routinely omitted; including them changes the
conclusion.

Second, ``live'' is not ``advantageous.'' The feedback reservoir does
transform a useless open-loop model into a predictive one. A study that
compared feedback QRC only against open-loop QRC, or against an untuned
ESN, would have reported a success. The advantage evaporates only against a
classical recurrent model tuned as hard as the quantum one.

Third, the scope is bounded and we state it as such. These are statevector
simulations at $q \le 11$, sizes a classical computer represents exactly, on
tasks with low-dimensional classical inputs. We make no claim about larger
systems, genuinely quantum-state inputs, or non-Clifford resources, where
provable learning separations are known to exist~\cite{huang2022} and the
relevant comparisons may behave differently. Our claim is the narrow, falsifiable
one our experiments support: at currently simulable scales, with fair baselines,
these two QRC advantage mechanisms do not beat their classical counterparts.

\section{Conclusion}

We have shown, under fair comparison, that two representative
quantum-reservoir advantage mechanisms, a two-point correlator readout and a
feedback-driven reservoir, do not outperform capacity- and tuning-matched
classical baselines on the tasks studied, in noiseless simulation at small
scale. The two controls behind this conclusion, a capacity-matched classical
model and a same-budget recurrent network inside a correctness-gated
deterministic harness, are inexpensive, decisive, and routinely omitted; we
offer them as a reusable template for honest QRC benchmarking rather than as a
verdict on quantum reservoir computing as a whole.

\section*{Data Availability}

The result files and the analysis scripts that generate every table and figure in
this paper are available at
\url{https://github.com/pandey-tushar/Quantum_Chaos_solver} (branch
\texttt{paper-7-cartography}). The repository includes the per-seed result JSONs
for Case~I and Case~II, the scripts (\texttt{mv\_correlator\_qrc.py},
\texttt{feedback\_qrc.py}), the figure script (\texttt{make\_figures.py}), and a
README mapping each table to the command that reproduces it. All results are
deterministic and reproduce bit-for-bit from the provided random seeds.

\section*{Acknowledgment}

The author used Claude (Anthropic) to assist with running and optimizing the
simulation code. All technical content, results, and conclusions are the
author's own.

\end{document}